# Scan-less confocal phase microscopy based on dual comb spectroscopy of two-dimensional-image-encoding optical frequency comb

Eiji Hase[1,2], Takeo Minamikawa[2,3], Shuji Miyamoto[1], Ryuji Ichikawa[1], Yi-Da Hsieh[2,3], Kyuki Shibuya[1,2], Yoshiaki Nakajima[2,4], Akifumi Asahara[2,4], Kaoru Minoshima[2,4], Yasuhiro Mizutani[2,5], Tetsuo Iwata[2,3], Hirotsugu Yamamoto[2,6], and Takeshi Yasui[2,3,*]

[1]Graduate School of Advanced Technology and Science, The Tokushima University, 2-1, Minami-Josanjima, Tokushima 770-8506, Japan

[2]JST, ERATO, MINOSHIMA Intelligent Optical Synthesizer Project, 2-1, Minami-Josanjima, Tokushima 770-8506, Japan

[3]Graduate School of Technology, Industrial and Social Sciences, The Tokushima University, 2-1, Minami-Josanjima, Tokushima 770-8506, Japan

[4]Graduate School of Informatics and Engineering, The University of Electro-Communications (UEC), 1-5-1 Chofugaoka, Chofu, Tokyo 182-8585, Japan

[5]Graduate School of Engineering, Osaka University, 2-1, Yamadaoka, Suita, Osaka 565-0871, Japan

[6]Center for Optical Research and Education, Utsunomiya University, 7-1-2, Yoto, Utsunomiya, Tochigi 321-8585, Japan

*Corresponding author's e-mail address: yasui.takeshi@tokushima-u.ac.jp

## Abstract (less than 150 words→149 words)

Confocal imaging and phase imaging are powerful tools in life science research and industrial inspection. To coherently link the two techniques with different depth resolutions, we introduce an optical frequency comb (OFC) to microscopy. Two-dimensional (2D) image pixels of a sample were encoded onto OFC modes via 2D spectral encoding, in which OFC acted as an optical carrier with a vast number of discrete frequency channels. Then, a scan-less full-field confocal image with a depth resolution of 62.4 µm was decoded from a mode-resolved OFC amplitude spectrum obtained by dual-comb spectroscopy. Furthermore, a phase image with a depth resolution of 13.7 nm was decoded from a mode-resolved OFC phase spectrum under the above confocality. The phase wrapping ambiguity can be removed by the match between the confocal depth resolution and the phase wrapping period. The proposed hybrid microscopy approach will be a powerful tool for a variety of applications.

Main text: approximately 3,000 words→467+2196+758=3421 words

**1. Introduction (less than 500 words→467 words)**

Optical microscopy has been widely used as a powerful tool in life science research and industrial inspection. Therefore, the development of optical microscopy has been a principal driving force in these fields. Among various types of optical microscopy, scanning confocal laser microscopy (CLM) [1-3] has attracted attention due to its three-dimensional (3D) imaging capability with confocality, enabling a confocal depth resolution of wavelength $\lambda$, spray light elimination, and a wide range of depths. Recently, scan-less CLM was achieved by a combination of one-dimensional (1D) spectral encoding CLM with line-field CLM [4] or two-dimensional (2D) spectral encoding [5, 6], enabling rapid image acquisition and robustness to external disturbances. If the depth resolution in these scan-less CLM images can be further decreased beyond the limit of the confocality, CLM will become an even more powerful tool in a variety of applications and fields.

Another type of optical microscopy with interesting features is digital holographic microscopy (DHM) [7-9], which enables 3D imaging with sub-wavelength (sub-$\lambda$) depth resolution using phase images reconstructed from a digital hologram. Although such 3D images can be acquired without the need for mechanical scanning, the range of depth is limited to less than the phase wrapping period $\lambda/2$ due to the phase wrapping ambiguity. Although the range of depth can be increased by the use of synthesized wavelengths between two laser lights at different wavelengths, it remains less than 10 µm [10, 11].

Thus, CLM and DHM both have advantages and disadvantages and are complementary to each other in depth measurement. If the confocality is given the phase image to match the confocal depth resolution with the phase wrapping period, a sub-$\lambda$ depth resolution can be achieved with a wide range of depths without the phase wrapping ambiguity. As such, scan-less confocal phase imaging will largely expand the applications of optical microscopy.

In this article, we applied an optical frequency comb (OFC) [12-14] to scan-less confocal phase imaging. Although OFCs have been widely used as an optical frequency ruler that is traceable to a frequency standard in optical frequency metrology, we propose another use of OFCs, namely an optical carrier having a vast number of discrete frequency channels with a frequency spacing equal to the repetition frequency $f_{rep}$. After encoding 2D image pixels of a sample on each OFC mode via the bidirectional conversion between wavelength and 2D space (namely, 2D spectral encoding) [5, 6, 15, 16], the 2D image can be decoded from a single mode-resolved OFC spectrum due to the one-to-one correspondence between image pixels and OFC modes. Fortunately, current state-of-the-art dual-comb spectroscopy, namely DCS, enables us to rapidly, precisely, and accurately acquire the mode-resolved OFC amplitude and phase spectra without the need for mechanical scanning [17-20]. We demonstrate a proof-of-principle experiment of scan-less confocal phase imaging by DCS of a 2D-image-encoding OFC.

**Results (2196 words)**

Principle of operation

Figure 1(a) illustrates the principle of scan-less CLM based on 2D spectral encoding [5, 6]. After passing through a light-source pinhole, each OFC mode is individually diffracted at different solid angles by a 2D spectral disperser for a wavelength-to-space conversion. The spatially dispersed OFC modes form a 2D spectrograph of the OFC and are then focused at different positions of a sample as a 2D array of focal spots by an objective lens combined with a relay lens (not shown). After being reflected by the sample, the 2D spectrograph is spatially overlapped again for the space-to-wavelength conversion. As such, the bidirectional conversion between the wavelength and 2D space modulates the spectral shape of the OFC depending on the 2D information of the sample. In this way, a one-to-one correspondence is established between the 2D image pixels and the OFC modes. The spectrally modulated OFC light passes through the detection pinhole, giving confocality. The mode-resolved amplitude and phase spectra of the 2D-image-encoded OFC are acquired by DCS. Finally, the confocal 2D image of the sample is decoded from the mode-resolved OFC spectra.

Figure 1(b) presents the principle of the confocal phase imaging. In CLM, only volume information in the vicinity of the focal point, namely the confocal volume (depth = $\Delta z$), can be obtained due to the confocality. DCS provides the mode-resolved amplitude and phase spectra for this confocal volume. The 2D amplitude image, decoded from the mode-resolved amplitude spectrum, gives the 2D mapping of reflection, absorption, or scattering of the sample for the entire confocal volume. Therefore, the 2D amplitude image is equivalent to the confocal 2D image acquired by CLM. In contrast, the 2D phase image, decoded from the mode-resolved phase spectrum, gives the 2D phase mapping in the confocal volume. While this 2D phase image is similar to that in DHM where sub-$\lambda$ depth resolution can be achieved [7-9], a clear difference between them is the existence of confocality. Therefore, we call it the confocal phase image. Here, if the confocal depth resolution $\Delta z$ is set equal to the phase wrapping period $\lambda/2$, the confocality can be coherently linked with the phase image. In other words, the confocality not only resolves the phase wrapping ambiguity but also achieves a wide range of depths, similar to CLM. Therefore, the use of both amplitude and phase images enables a sub-$\lambda$-resolved phase image with the confocality and a $\lambda$-resolved confocal image with a wide range of depths.

Experimental setup

Figure 1(c) is a schematic diagram of the experimental setup, including a dual femtosecond Er-fibre OFC (OFC1 or signal comb, $f_{ceo1}$ = 21.4 MHz, $f_{rep1}$ = 100,387,960 Hz; OFC2 or local comb, $f_{ceo2}$ = 21.4 MHz, $f_{rep2}$ = $f_{rep1}$ - $\Delta f_{rep}$ = 100,389,194 Hz; $\Delta f_{rep}$ = $f_{rep2}$ - $f_{rep1}$ = 1,234 Hz), a 2D spectral disperser [5, 6, 15, 16] composed of a virtually imaged phased array (VIPA) [21] and a diffraction grating, relay lenses, an objective lens, a confocal pinhole, and DCS optical systems. The OFC1 (signal comb) light forms a 2D array of focal spots, corresponding to a 2D spectrograph of the OFC1 light, on a sample, by passing through the 2D spectral disperser, the relay lenses, and the objective lens. The reflected OFC1 light inversely propagates the same optics for the spatial overlapping of each OFC mode, resulting in encoding of a 2D image on the OFC1 spectrum. After passing through the detection pinhole for confocality, the OFC1 light interferes with the OFC2 (local comb) light to generate an interferogram. Finally, the mode-resolved amplitude and phase spectra for the OFC1 light are obtained by Fourier transform of the acquired interferogram.

Scan-less confocal phase imaging

We first measured the mode-resolved amplitude and phase spectra when a 1951 USAF resolution test chart (positive type) was placed at the focal position (d = 0 μm) as a sample. We accumulated 100 temporal waveforms of interferogram acquired at $\Delta f_{rep}$ (= 1,234 Hz) and performed Fourier transform of it. Figures 2(a) and 2(b) show the mode-resolved amplitude and phase spectra obtained in no pattern area of the sample, whereas Figs. 2(c) and 2(d) show the spectra obtained in the pattern area (data acquisition time = 81 ms). The fine structure inside the spectra in Figs. 2(a) and 2(c) reflects the transmission spectrum of VIPA, and the phase noise increased at frequencies with low transmission in Figs. 2(b) and 2(d). The difference between the spectra with and without the pattern was caused by a 2D distribution of reflectivity or phase change in the sample. For example, the modulation of spectral envelope in Fig. 2(c) and more reddish part of the spectra in Fig. 2(d) reflect line patterns in the test chart. In this way, we obtained the 2D-image-encoded, mode-resolved OFC1 spectra.

We next decoded a 2D amplitude image of the test chart from the mode-resolved amplitude spectra in Fig. 2(c). To eliminate the influence of the spectral shape in the OFC1 light, a normalized amplitude spectrum was obtained by using the spectrum in Fig. 2(a) as a reference. Figure 3(a) shows a 2D amplitude image of the test chart (image size = 760 × 168 μm, pixel size = 82 × 151 pixels, image acquisition time = 81 ms) when the sample was placed at the focal position (d = 0 μm). This image is corresponding to the confocal image [see Fig. 1(b)]. A pattern of Element 1 in Group 4 (spatial frequency = 16 lp/mm) was confirmed clearly. Although the image distortion in the 2D spectral encoding was compensated when decoding the image from the spectrum, it was somewhat remained. To confirm the confocality, we move the position of the test chart before, at, and after the focal position. Figure 3(b) shows the 2D amplitude image when the test chart was out of focus (d = +100 μm) whereas an upper panel of Movie 1 shows a series of 2D amplitude images when the sample was moved along a direction vertical to the chart surface (d = -215 μm ~ +215 μm). The amplitude image was changed depending on the depth position due to the confocality. In this way, scan-less full-field confocal imaging was successfully demonstrated.

We further decoded a 2D phase image of the test chart from the mode-resolved phase spectrum [see Fig. 2(d)]. To eliminate the influence of the initial phase in the OFC1 light, we subtracted the reference phase spectrum [see Fig. 2(b)] from the signal phase spectrum. Figure 3(c) shows a 2D phase image (image size = 760 × 168 μm, pixel size = 82 × 151 pixels, image acquisition time = 81 ms) when we set the test chart in focus (d = 0 μm). This is corresponding to the confocal phase image in Fig. 1(b). Although the obtained phase image was similar to the amplitude image in Fig. 3(a), its image contrast arises from the phase difference between the pattern region and the surrounding region, which are corresponding to reflection-coated region and the uncoated region, respectively. The thickness of the reflection-coated layer was determined to be 116±7.4 nm from the phase difference of 0.94±0.06 rad. This thickness was in good agreement with the surface unevenness (= 122.7±1.8 nm) measured by an atomic force microscope (Olympus Corp., OLS3500-PTU, depth resolution = 1.0 nm). When the test chart was moved to out of focus, the image contrast significantly decreased due to the confocality although the phase offset was changed depending on the depth position, as shown in Fig. 3(d) and the right panel of Movie 1. From these results, it can be seen that the proposed system accurately acquired a 3D image of the sample with sub-$\lambda$ depth resolution under the confocality.

To highlight the fast imaging capability in the proposed system, we acquired a series of confocal amplitude and phase images of the test chart in motion along in-plane direction without signal accumulation. Figures 3(e) and 3(f) shows a single shot image of them whereas Movie 2 shows a series of confocal amplitude and phase images. The acquisition time of the single shot image was only 810 µs and the frame rate in Movie 2 was achieved to 1,234 Hz (= $\Delta f_{rep}$). Although the obtained amplitude and phase images have less contrast than those in Figs. 3(a) and 3(c) or Movie 1 due to the decreased image acquisition time, the state of the moving test chart was visualized in the confocal amplitude and phase image.

Finally, we demonstrated 3D shape measurement based on the confocal phase image. As a sample, we made a three-step structure with nm order on a silicon substrate, whose surface was covered by a gold thin film. Figures 4(a) and 4(b) show the 2D amplitude image and the 2D phase image of this sample (image size = 671 × 147 µm, pixel size = 72 × 132 pixels, image acquisition time = 81 ms). The amplitude image indicates the step structure with low contrast because of little difference of reflectivity among different steps. On the other hand, the phase image clearly visualizes the step structure due to the phase-based image contrast. Figure 4(c) shows the 3D shape of the sample calculated from the phase image in Fig. 4(b). We determined the step difference to be $86.0 \pm 13.6$ nm for the 1st/2nd step-difference and $139.9 \pm 14.3$ nm for the 2nd/3rd step-difference. These values were in good agreement with them measured by the atomic force microscope: 86.0±0.1 nm for the 1st/2nd step-difference and 140±0.3 nm for the 2nd/3rd step-difference.

Evaluation of spatial resolution

We next considered the spatial resolution of the proposed system. A detailed spatial distribution of the 2D-spectral-encoding OFC1 light on a sample is given in the Methods. Figure 5(a) is a schematic drawing of a 2D array of focal spots on a sample when the transmission resonance spectrum of the VIPA includes a single OFC1 mode [16]. In contrast, if multiple modes of OFC1 are contained in the transmission resonance spectrum of the VIPA, a series of focal spots is spatially overlapped along the vertical direction and is still discrete along the horizontal direction, as shown in Fig. 5(b) [5, 6]. In the present setup with a VIPA resonance linewidth = 136 MHz, VIPA free spectral range (FSR) = 15.1 GHz, OFC1 optical bandwidth = 1.24 THz, and $f_{rep1}$ = 100 MHz, the 2D array of 82 spots by 151 spots was distributed within an image size of 760 µm by 168 µm, as shown in Fig. 5(b). Because the spatial resolution for this condition is similar to that in the scan-less CLM based on 2D-image-encoding [5, 6], the spatial resolution is respectively estimated to be 4.7 µm for the horizontal dimension and 2.0 µm for the vertical dimension from the spectral resolution of the grating and VIPA combined with the OL. We determined the actual spatial resolution from the 2D amplitude image of a knife edge to be 9.28 µm for the horizontal direction and 1.61±0.13 µm for the vertical direction. The former resolution was limited by discrete distribution of horizontal spots whereas the latter resolution was limited by the instrumental resolution of the 2D spectral disperser. These values are in reasonable agreement with the expected values.

Although one-to-one correspondence could not be established between the image pixels and OFC modes in the present setup, this condition did not increase the spatial resolution because the instrumental resolution is larger than the comb-mode linewidth and the comb spacing in OFC1. It is not very difficult to achieve one-to-one correspondence in the form of Fig. 5(a) using a commercially available OFC with a higher $f_{rep}$ (for example, $f_{rep}$ = 250 MHz). The discrete 2D distribution of focal spots in Fig. 5(a) has two benefits for imaging. First, crosstalk between adjacent pixels can be suppressed for both the horizontal and vertical dimensions. Second, image deconvolution [22] can be applied by the use of spectral interleaving in OFC [23-25]. These benefits can be achieved only using an OFC with discrete channels as a light source.

We next considered the depth resolution of the proposed system. The depth resolution in the 2D amplitude image is limited by the confocality in this system, in the same manner as traditional CLM [3]. We confirmed the confocal depth resolution of 61.4 µm in the present system (not shown), which is in reasonable agreement with the theoretical value (=43.8 µm). Although this confocal depth resolution was still larger than the phase wrapping period $\lambda/2$ (= 0.775 µm) in the present setup, it will be possible to match the period by using tight confocal optics and higher-NA OL. In contrast, the depth resolution in the 2D phase image is limited by the phase noise of the signal in the same manner as DHM. Because the phase noise in the present system was approximately $\lambda/226$, the depth resolution was estimated to be 6.9 nm, which is reasonably consistent with the step measurement in the test chart and the three-step-structure sample. If the confocal depth resolution is set equal to the phase wrapping period $\lambda/2$, a depth resolution of 13.7 nm and a range of depth up to the working distance of the OL (≈ several mm) can be achieved simultaneously without the phase wrapping ambiguity.

**Discussion (758 words)**

We demonstrated that full use of the mode-resolved amplitude and phase spectra enables the fusion of confocal imaging and phase imaging as complementary to each other; to the best of our knowledge, this is the first demonstration of this fusion. Although it is well known that DCS provides both amplitude and phase spectra due to Fourier transform spectroscopy, these two full spectra have not been so actively used in previous applications of DCS. Recently, a similar approach was effectively applied for polarization-modulation-free DCS ellipsometry [26]. These demonstrations clearly indicate that simultaneous use of amplitude and phase spectra has the potential to enhance the utility of DCS and may lead to novel applications of DCS.

One may consider the similarity of the proposed method to 2D-spectral-encoding scan-less CLM [5, 6]. Although 2D-spectral-encoding was used for the scan-less measurement in both methods, the proposed method has several significant differences from 2D-spectral-encoding CLM. The main difference is the simultaneous acquisition of the confocal image and phase image as demonstrated above. Another difference is the ultra-discrete sampling in the OFC mode. Such ultra-discrete sampling is beneficial because of the improved signal-to-noise ratio (SNR) in this imaging method in addition to the reduction of pixel crosstalk and the adoption of image deconvolution. The OFC has a discretely localized distribution of optical energy with a constant frequency spacing in the optical frequency region, and DCS enables one to acquire the entire optical energy in the radio-frequency (RF) region without loss, as shown in Fig. 6(a). In contrast, the noise component is not localized and is continuously distributed throughout the entire RF region. The acquisition of the mode-resolved OFC spectrum by DCS enables us to select only the signal components and a few noise components around the comb modes while rejecting the remaining noise components existing in the frequency gap between comb modes. The noise rejection efficiency depends on the ratio of the spectral comb spacing to the comb-mode linewidth in DCS (typically, >10), and the SNR should be greatly enhanced at the position of the comb modes. In contrast, in 2D-spectral-encoding CLM, a broadband continuous spectrum is acquired by an optical spectrum analyser. Because neither the signal nor the noise components are localized but are continuously distributed, as shown in Fig. 6(b), all noise components contribute to the SNR; in other words, there is no noise rejection. If the total optical energy and the total noise component—the entire area of the red and blue regions shown in Figs. 6(a) and 6(b)—are equivalent to each other in these two methods, the SNR in Fig. 6(a) should be better than that in Fig. 6(b). Such an enhanced SNR will directly lead to enhanced image contrast or decreased image acquisition time. Unfortunately, the limited dynamic range of the photodetector used in DCS blurs these effects in the presented results. However, if the optical comb spectrum is optimized to the limited dynamic range by spectral shaping [27] or mode filtering [28], the enhanced SNR obtained by noise rejection will be clearer.

Finally, we discuss the versatility of the proposed method. In this article, we achieved hybrid confocal and phase imaging under scan-less conditions by imposing 2D image pixels on an OFC mode via 2D spectral encoding. If another physical quantity to be measured is superimposed on each OFC mode by another dimensional conversion, a vast amount of data for the measured quantity can be simultaneously obtained from a single mode-resolved OFC spectrum. Such an approach, which we refer to as dimensional-conversion OFC, will offer the possibility of expanding the application fields of OFC beyond optical frequency metrology because this method can be applied for other optical metrologies by using a variety of dimensional conversions. For example, time-to-wavelength conversion [29] and polarization-to-wavelength conversion [30] are promising methods for dimensional-conversion OFC. The ultra-narrow OFC mode linewidth enables an infinitesimal resolution of the physical quantity, whereas an ultra-constant sampling interval can be achieved by the frequency spacing of the OFC. An overly discrete sampling interval can be reduced by use of a spectral interleaving OFC [23-25]. Therefore, dimensional-conversion OFC has the potential to exceed the limit in the direct measurement of the physical quantity.

In summary, we have successfully demonstrated scan-less confocal phase imaging by DCS of a 2D-image-encoding OFC. Both a confocal amplitude image with a depth resolution of 62.4 μm and a confocal phase image with a depth resolution of 13.7 nm were decoded from the mode-resolved amplitude and phase spectra in a 2D-image-encoding OFC. The proposed hybrid microscopy will boost life science research and industrial inspection.

**Methods (less than 3,000 words→883 words)**

Experimental setup

Figure 1(c) shows a schematic diagram of the experimental setup. We used a home-made femtosecond Er-fibre OFC (OFC1 or signal comb; centre wavelength = 1555 nm, mean output power = 80 mW, $f_{ceo1}$ = 21.4 MHz, $f_{rep1}$ = 100,387,960 Hz) for 2D spectral encoding. $f_{ceo1}$ and $f_{rep1}$ were phase-locked to a rubidium frequency standard (Stanford Research Systems, FS725, accuracy = 5 × $10^{-11}$, instability = 2 × $10^{-11}$ at 1 s, not shown) via a laser control system. After passing through an optical bandpass filter (passband = 1538~1562 nm, not shown), the OFC1 light passed through a beam splitter and was then fed into an optical system for 2D spectral encoding, which is composed of a 2D spectral disperser, a pair of relay lenses, and an objective lens. The 2D spectral disperser is composed of a virtually imaged phased array (VIPA; Light Machinery, OP-6721-6743-8, free spectral range = 15.1 GHz, finesse = 110) and diffraction grating (Spectrogon, PC 1200 30 × 30 × 6, groove density = 1200 grooves/mm, efficiency = 90 %), enabling it to spread OFC1 modes in 2D space, known as a 2D spectrograph of OFC1. The 2D spectrograph was relayed to the entrance pupil of an objective lens (dry, numerical aperture = 0.25, working distance = 16 mm) by a pair of relay lenses at different solid angles depending on its wavelength and was then focused onto a sample with the objective lens. This approach resulted in the formation of a 2D focal spot array of the 2D spectrograph. The 2D image information of the sample was encoded onto the 2D spectrograph via the reflection, absorption, scattering, or phase change. As the encoded 2D spectrograph inversely passed through the same optical system, it was spatially overlapped again as a 2D-image-encoded OFC1 light. After being reflected by the beam splitter and passing through a detection pinhole (diameter = 25 μm) equipped with a pair of lenses (focal length = 50 mm), the 2D-image-encoded OFC1 light was fed into the experimental setup of the DCS. We used another femtosecond Er-fibre OFC (OFC2 or local comb; centre wavelength= 1555 nm, mean output power = 80 mW, $f_{ceo2}$ = 21.4 MHz, $f_{rep2}$ = 100,389,194 Hz, $\Delta f_{rep}$ = $f_{rep2}$ - $f_{rep1}$ = 1,234 Hz) for a local comb in DCS. The local comb with an intra-cavity electro-optical modulator was tightly and coherently locked to the signal comb with a certain frequency offset $\Delta f_{rep}$ using a narrow-linewidth CW laser (CW-ECLD, Redfern Integrated Optics Inc., PLANEX, centre wavelength: 1,550 nm; FWHM: <2.0 kHz) as an intermediate laser [31, 32]; this coherent locking allows us to perform coherent averaging (accumulation) of the interferograms observed in DCS [33, 34]. The 2D-image-encoded OFC1 light interfered with the OFC2 light by spatially overlapping with a polarization beam splitter. The generated interferogram was detected by a combination of a half-wave plate, another polarization beam spliter, and a balanced detector (not shown). The detected electric signal was acquired within a time window of 81 µs with a digitizer (National Instruments Corp., NI PCI-5122, sampling rate = $f_{rep2}$ = 100,389,194 samples/s, number of sampling points = 81,353, resolution = 14 bit). The acquire signal is corresponding to the interferogram signal with a time window of 9.96 ns at a sampling interval of 122 fs. Finally, the mode-resolved OFC1 amplitude and phase spectra were obtained by Fourier transform of the acquired interferogram and were used to decode the 2D image information of the sample.

Spatial distribution of focal spots based on 2D spectral encoding

Figure 5(a) shows a schematic drawing for the 2D array of focal spots on a sample when the transmission resonance spectrum of the VIPA (linewidth = $\Delta f_{VIPA}$) includes a single OFC1 mode (namely, $\Delta f_{VIPA} < f_{rep1}$) [16]. Because the VIPA disperses the OFC1 spectrum along the vertical direction while the following grating disperses the spectrum horizontally, the OFC modes with the lowest optical frequency (longest wavelength) to the highest optical frequency (shortest wavelength) are discretely developed in 2D space along the dashed line, enabling a one-to-one correspondence between image pixels and OFC modes. The horizontal dimension of this 2D distribution corresponds to the ratio of the optical bandwidth of OFC1 (= $\Delta \nu_{OFC1}$) to the FSR of the VIPA (= $FSR_{VIPA}$), and the vertical dimension depends on $FSR_{VIPA}$. The spacing between horizontally adjacent spots is determined by $FSR_{VIPA}$ whereas that between vertically adjacent spots is determined by $f_{rep1}$. Although the horizontal and vertical dimensions of the focal spot should depend on the linewidth of each OFC mode (= $\Delta\nu_{mode}$) under an infinite spectral resolution, the actual dimensions were respectively limited by the finite spectral resolution of the 2D spectral disperser [5, 6].

In contrast, if multiple modes of OFC1 are contained in the transmission resonance spectrum of the VIPA ($\Delta f_{VIPA} < f_{rep1}$), a series of focal spots along the vertical direction is spatially overlapped although the series is still discrete along the horizontal direction, as shown in Fig. 5(b). In other words, a one-to-one correspondence was established between the image pixels and the OFC mode only along the horizontal direction. Because the horizontal and vertical dimensions of the focal spot in Fig. 5(b) are similar to those in Fig. 5(a), the only difference between the results is whether the image pixels along the vertical direction are continuous or discrete.

**Acknowledgements**

This work was supported by grants for the Exploratory Research for Advanced Technology (ERATO) MINOSHIMA Intelligent Optical Synthesizer (IOS) Project (JPMJER1304) from the Japanese Science and Technology Agency and a Grant-in-Aid for Scientific Research (A) No. 26246031 and 15H02026 from the Ministry of Education, Culture, Sports, Science, and Technology of Japan. The authors acknowledge Dr. Toshihiro Okamoto, Dr. Takahiko Mizuno, and Mr. Shun Kamada of The Tokushima University, Japan for their help in the sample preparation and evaluation. The authors also acknowledge Ms. Shoko Lewis and Ms. Natsuko Takeichi of Tokushima Univ., Japan for her help in the preparation of the manuscript.## Author contributions

T. Y., H. Y., and K. M. conceived the project. E. H., S. M., R. I., Y.-D. H., and K. S. performed the experiments and/or analysed the data. Y. N. and A. A. contributed to the dual-comb sources. E. H. and T. Y. wrote the manuscript. T. M., Y. M., and T. I discussed the results and commented on the manuscript.

## Competing financial interests statement

The authors declare no competing financial interests.

-20-

# References

[1] Davidovits, P. & Egger, M. D. Photomicrography of corneal endothelial cells *in vivo*. *Nature* **244**, 366-367 (1973).

[2] Brakenhoff, G. J., Blom, P., & Barends, P. Confocal scanning light microscopy with high aperture immersion lenses. *J. Microsc.* **117**, 219-232 (1979).

[3] Sheppard, C. J. & Shotton, D. M. *Confocal laser scanning microscopy* (BIOS Scientific Publishers, Oxford, 1997).

[4] Kim, J., Kang, D., & Gweon, D. Spectrally encoded slit confocal microscopy. *Opt Lett.* **31**, 1687-1689 (2006).

[5] Tsia, K. *et al*. Simultaneous mechanical-scan-free confocal microscopy and laser microsurgery. *Opt. Lett.* **34**, 2099-2101 (2009).

[6] Tsia, K. K., Goda, K., Capewell, D., & Jalali, B. Performance of serial time-encoded amplified microscope. *Opt. Express* **18**, 10016-10028 (2010).

[7] Marquet, P. *et al.* Digital holographic microscopy: a noninvasive contrast imaging technique allowing quantitative visualization of living cells with subwavelength axial accuracy. *Opt. Lett.* **30**, 468-470 (2005).

[8] Kemper, B. & Bally, G. Digital holographic microscopy for live cell applications and technical inspection. *Appl. Opt.* **47**, A52-A61 (2008).

[9] Kim, M. K. *Digital holographic microscopy* (Springer New York, New York, 2011).

[10] Gass, J., Dakoff, A., & Kim, M. K. Phase imaging without $2\pi$ ambiguity by

multiwavelength digital holography. *Opt. Lett.* **28**, 1141-1143 (2003).

[11] Myung, D. & Kim, K. Digital holographic microscopy with dual-wavelength phase unwrapping. *Appl. Opt.* **45**, 451-459 (2006).

[12] Udem, Th., Reichert, J., Holzwarth, R., & Hänsch, T. W. Accurate measurement of large optical frequency differences with a mode-locked laser. *Opt. Lett.* **24**, 881-883 (1999).

[13] Niering, M. *et al.* Measurement of the hydrogen 1S-2S transition frequency by phase coherent comparison with a microwave cesium fountain clock. *Phys. Rev. Lett.* **84**, 5496-5499 (2000).

[14] Udem, Th., Holzwarth, R., and Hänsch, T. W. Optical frequency metrology. *Nature* **416**, 233-237 (2002).

[15] Xiao, S. & Weiner, A. M. 2-D wavelength demultiplexer with potential for ≥ 1000 channels in the C-band. *Opt. Express* **12**, 2895-2902 (2004).

[16] Diddams, S. A., Hollberg, L., & Mbele, V. Molecular fingerprinting with the resolved modes of a femtosecond laser frequency comb. *Nature* **445**, 627-630 (2007).

[17] Schiller, S. Spectrometry with frequency combs. *Opt. Lett.* **27**, 766–768 (2002).

[18] Keilmann, F., Gohle, C., & Holzwarth, R. Time-domain mid-infrared frequency-comb spectrometer. *Opt. Lett.* **29**, 1542–1544 (2004).

[19] Yasui, T. *et al.* Terahertz frequency comb by multifrequency-heterodyning photoconductive detection for high-accuracy, high-resolution terahertz spectroscopy.

*Appl. Phys. Lett.* **88**, 241104 (2006).

[20] Coddington, I., Newbury, N., & Swann, W. Dual-comb spectroscopy. *Optica* **3**, 414-426 (2016).

[21] Shirasaki, M. Large angular dispersion by a virtually imaged phased array and its application to a wavelength demultiplexer. *Opt. Lett.* **21**, 366–368 (1996).

[22] Kundur, D. & Hatzinakos, D. Blind image deconvolution. *IEEE Signal Process. Mag.* **13**, 43-64 (1996)

[23] Hsieh, Y.-D., *et al.* Spectrally interleaved, comb-mode-resolved spectroscopy using swept dual terahertz combs. *Sci. Reports* **4**, 3816 (2014).

[24] Yasui, T., *et al.* Super-resolution discrete Fourier transform spectroscopy beyond time window size limitation using precisely periodic pulsed radiation. *Optica* **2**, 460-467 (2015).

[25] Okubo, S., *et al.* Near-infrared broadband dual-frequency-comb spectroscopy with a resolution beyond the Fourier limit determined by the observation time windows. *Opt. Express* **23**, 33184-33193 (2015).

[26] Minamikawa, T., *et al.* Dual-optical-comb spectroscopic ellipsometry. *Technical Digest of Conference on Lasers and Electro-Optics 2016*, SW1H.5 (2016).

[27] Jiang, Z., Leaird, D. E., & Weiner A.M. Line-by-line pulse shaping control for optical arbitrary waveform generation. *Opt. Express* **13**, 10431-10439 (2005).

[28] Steinmetz, T., *et al.* Fabry–Pérot filter cavities for wide-spaced frequency combs with large spectral bandwidth. *Appl. Phys. B* **96**, 251–256 (2009).

[29] Minoshima, K., Matsumoto, H., Zhang, Z., & Yagi, T. Simultaneous 3-D imaging using chirped ultrashort optical pulses. *Jpn J. Appl. Phys.* 33, L1348-L1351 (1994).

[30] Oka, K. & Kato, T. Spectroscopic polarimetry with a channeled spectrum. *Opt. Lett.* **24**, 1475-1477 (1999).

[31] Nishiyama, A. *et al.* Doppler-free dual-comb spectroscopy of Rb using optical-optical double resonance technique. *Opt. Express* **24**, 25894-25904 (2016).

[32] Asahara, A. *et al.* Dual-comb spectroscopy for rapid characterization of complex optical properties of solids. *Opt. Lett.* **41**, 4971-4974 (2016).

[33] Baumann, E. *et al.* Spectroscopy of the methane $\nu_3$ band with an accurate midinfrared coherent dual-comb spectrometer. *Phys. Rev. A* **84**, 062513 (2011).

[34] Roy, J. Deschênes, J.-D. Potvin, S. & Genest, J. Continuous real-time correction and averaging for frequency comb interferometry. *Opt. Express* **20**, 21932-21939 (2012).

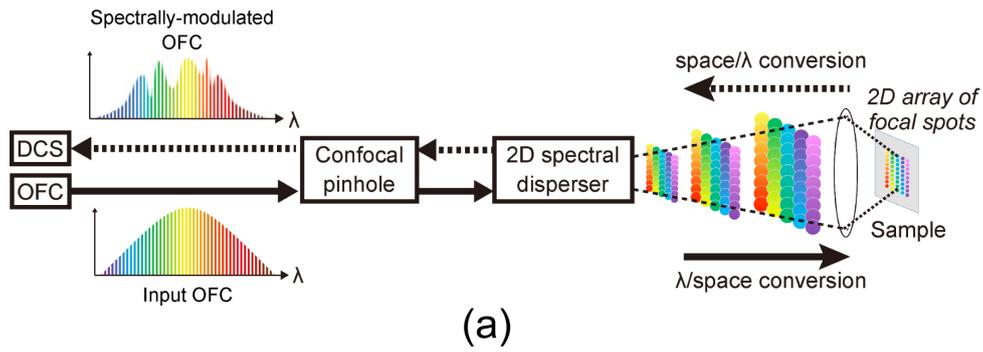

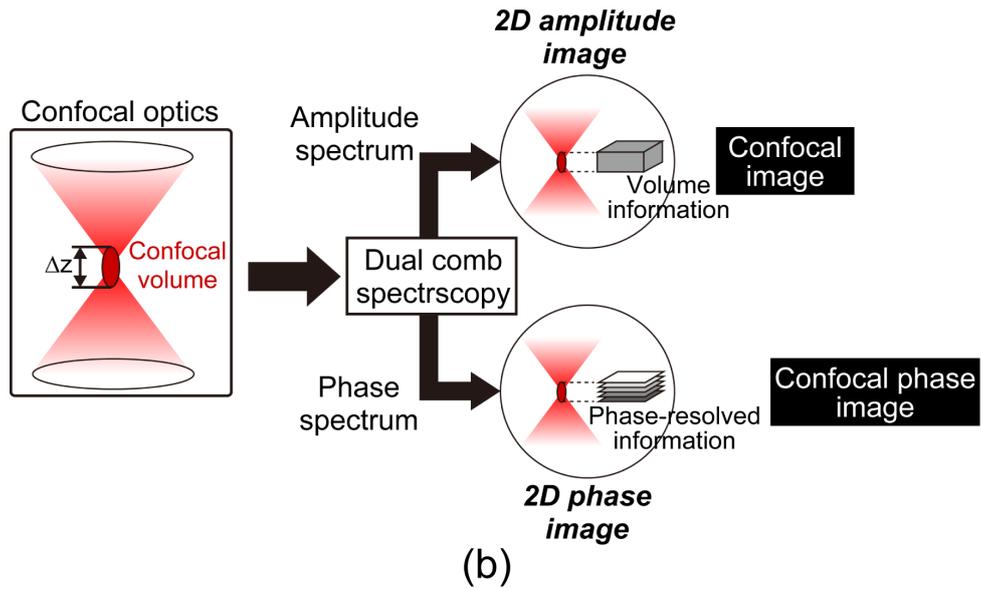

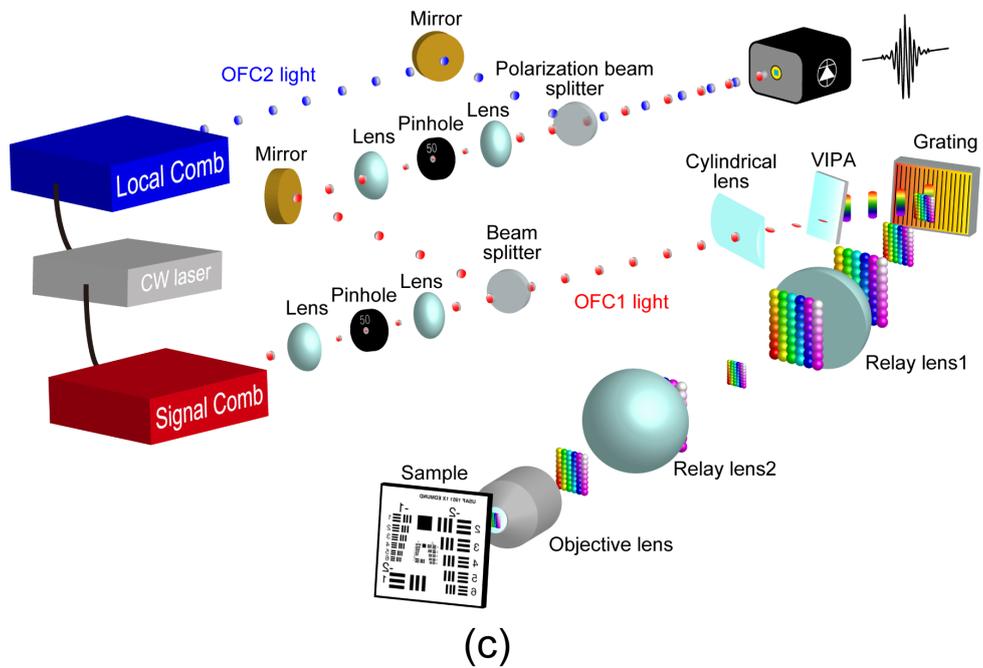

Fig. 1. Principle of operation: (a) scan-less CLM based on 2D spectral encoding and (b) confocal phase imaging. (c) Experimental setup.

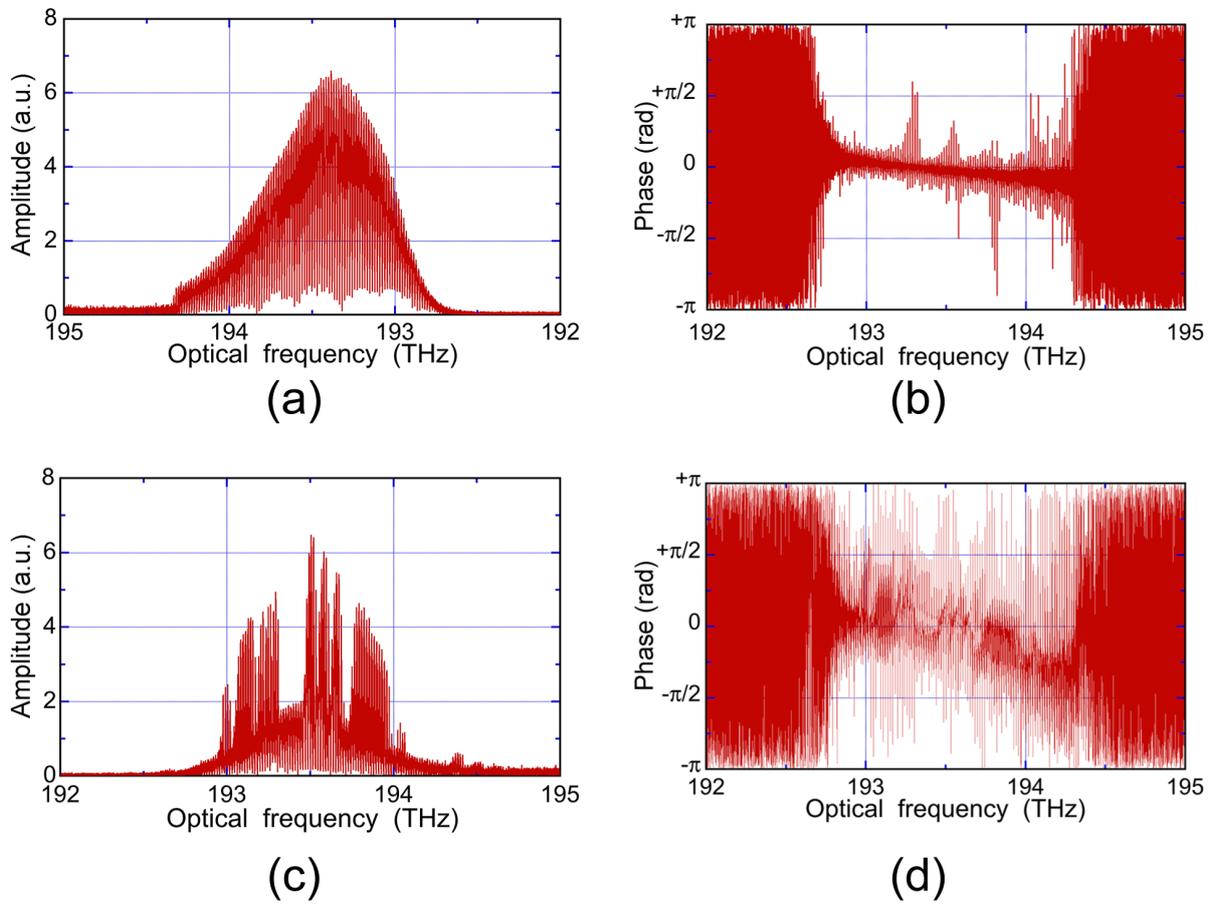

Fig. 2. Mode-resolved (a) amplitude and (b) phase spectra obtained in no pattern area. Mode-resolved (c) amplitude and (d) phase spectra obtained in the pattern area. Data acquisition time is 81 ms

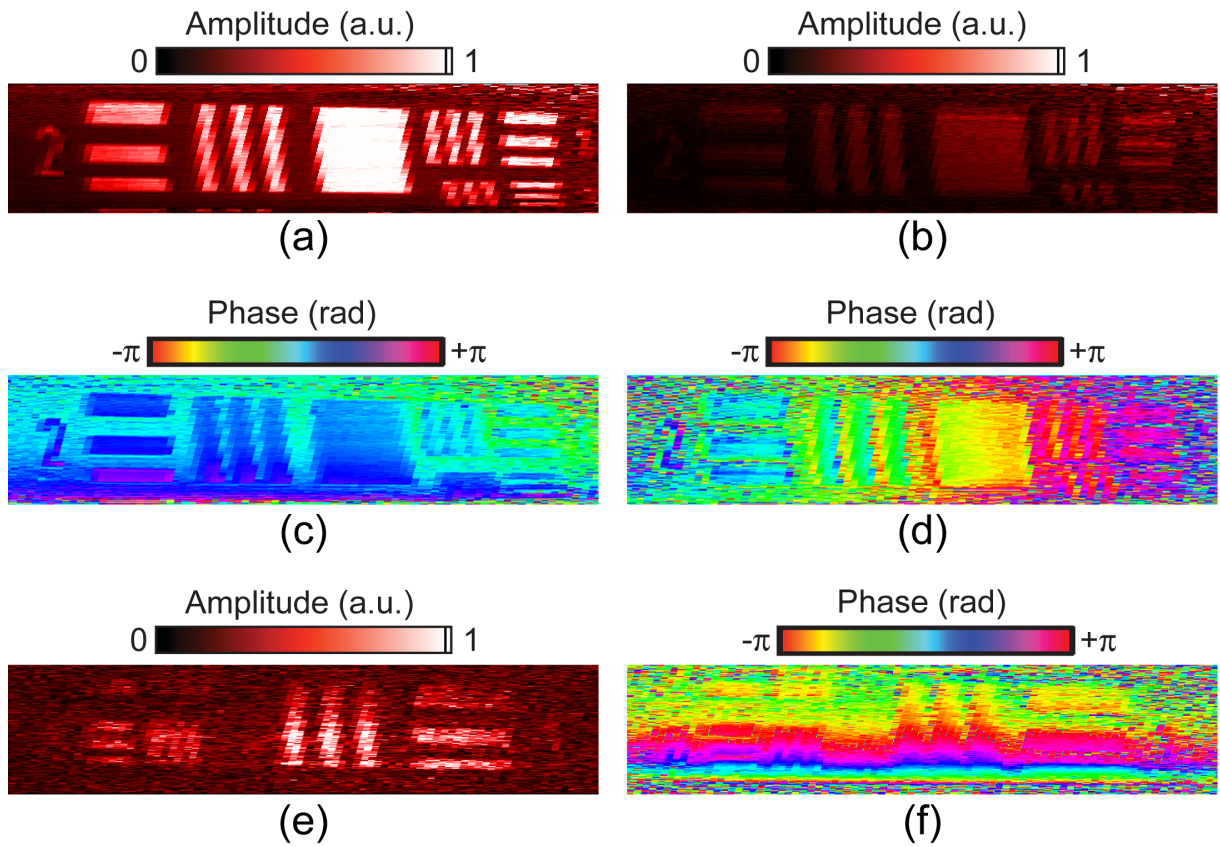

Fig. 3. Confocal amplitude image of a 1951 USAF resolution test chart with positive pattern when the sample was placed at (a) the focal position (d = 0 μm) and (b) out of focus (d = +100 μm). Confocal phase image of the test chart when the sample was placed at (c) the focal position (d = 0 μm) and (d) out of focus (d = +100 μm). Number of signal integration is 100, corresponding to the image acquisition time of 81 ms. Confocal (e) amplitude and (f) phase images of the test chart without signal integration (image acquisition time = 810 μs).

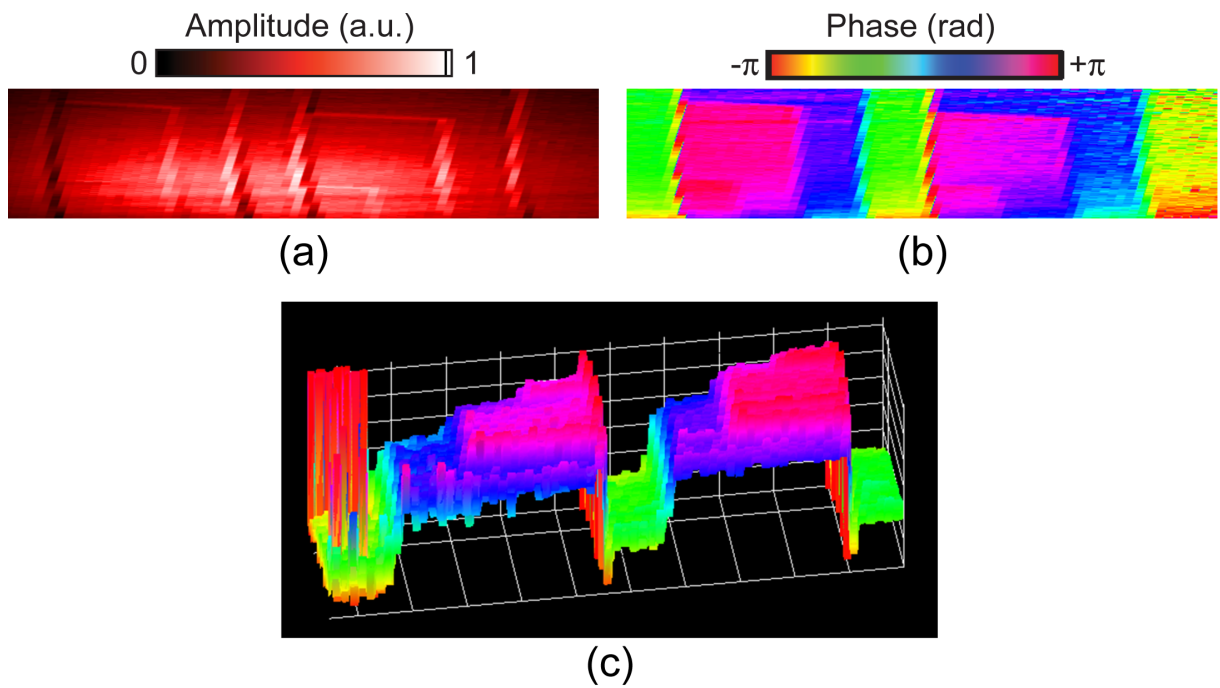

Fig. 4. (a) Confocal amplitude image and (b) confocal phase image of a three-step structure with nm order on a silicon substrate (image size = 760 × 168 μm, pixel size = 82 × 151 pixels, image acquisition time = 81 ms). (c) 3D shape of the three-step structure.

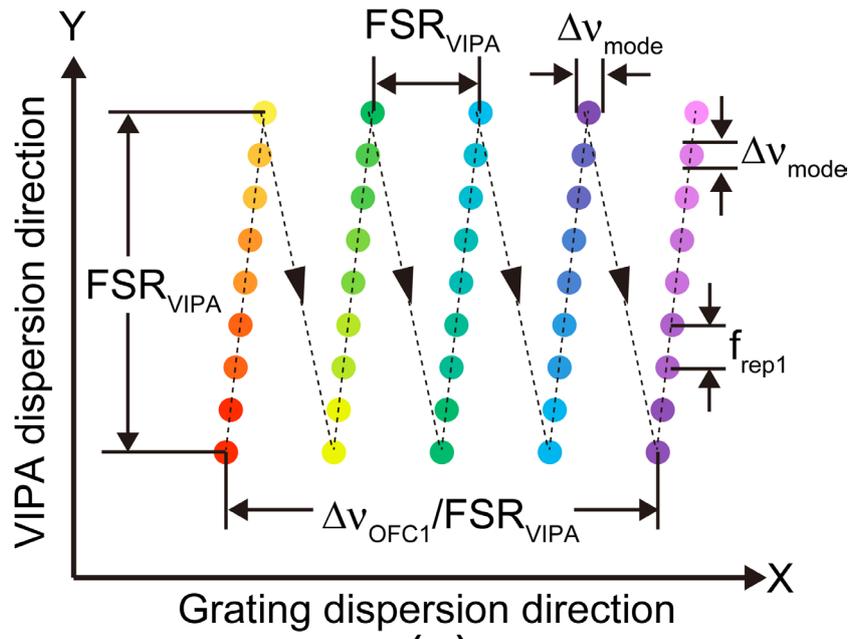

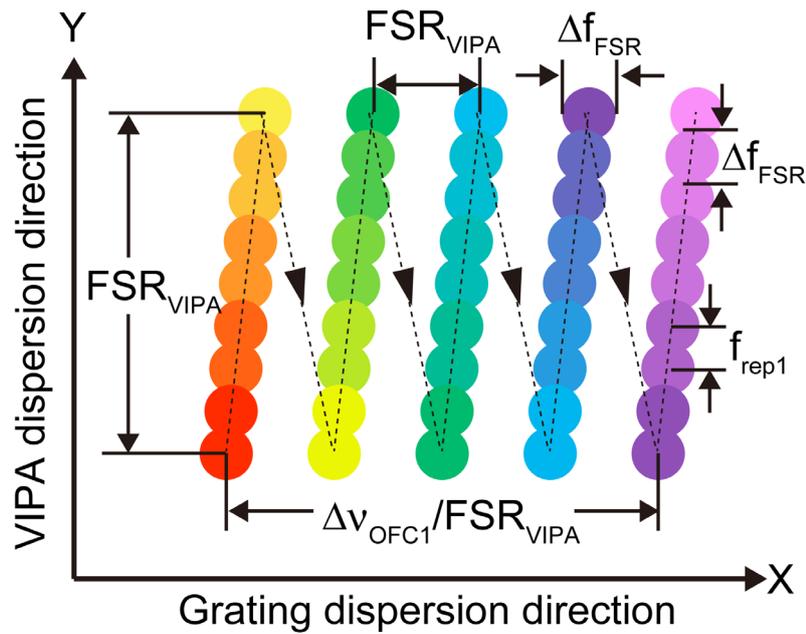

Fig. 5. Schematic drawing for 2D array of focal spots on a sample (a) when the transmission resonance spectrum of VIPA includes a single OFC1 mode and (b) when multiple modes of OFC1 are contained in the transmission resonance spectrum of VIPA.

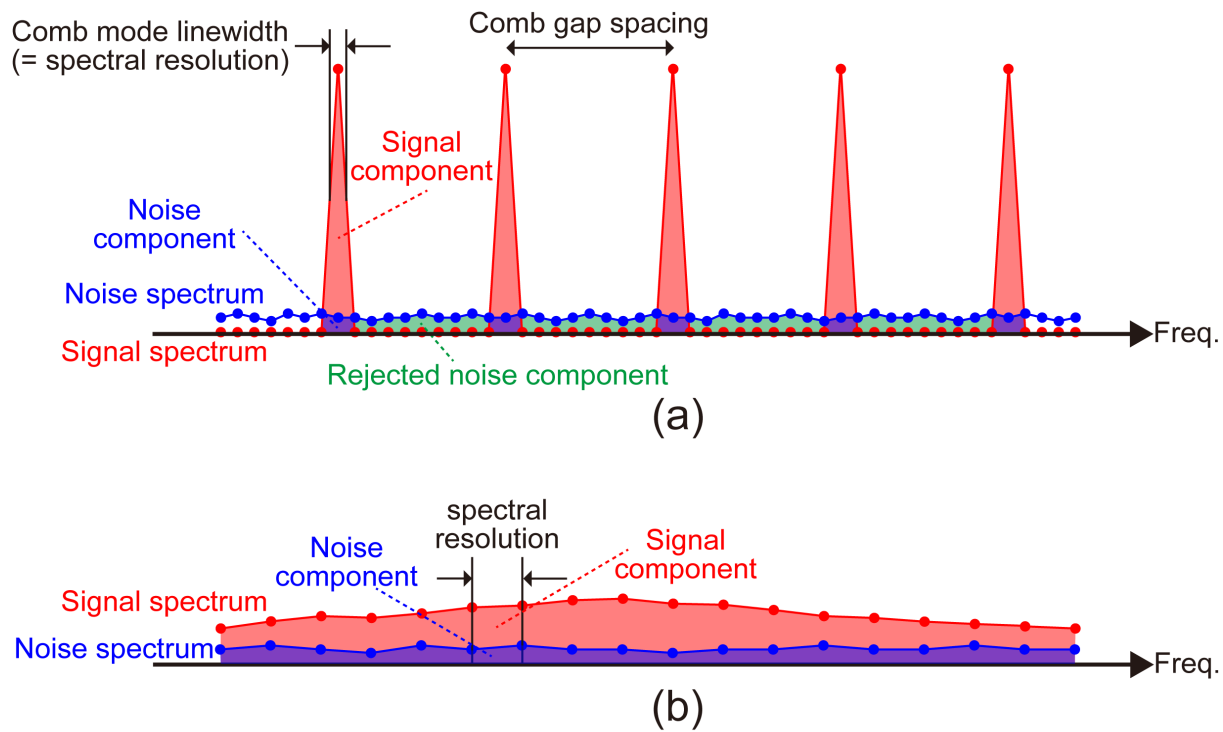

Fig. 6. Contributions of signal and noise in (a) discrete data acquisition and (b) continuous data acquisition.

Movie 1. A series of confocal amplitude and phase images when a 1951 USAF resolution test chart was moved along a direction vertical to the chart surface (image size = 760 × 168 μm, pixel size = 82 × 151 pixels, image acquisition time = 81 ms).

Movie 2. A series of confocal amplitude and phase images when a 1951 USAF resolution test chart was moved along in-plane direction (image size = 760 × 168 μm, pixel size = 82 × 151 pixels, frame rate = 1,234 frame/s).